**The essential role of multi-point measurements in investigations of heliospheric turbulence, three-dimensional structure, and dynamics**

W. H. Matthaeus[1]

Department of Physics and Astronomy, University of Delaware, Newark DE 19716 USA
email: whm@udel.edu   Tel: (302)-983-8767

Co-authors:

S. Adhikari[1], R. Bandyopadhyay[22], M. R. Brown[2], R. Bruno[29], J. Borovsky[3], D. Caprioli[5], V. Carbone[4], A. Chasapis[14], R. Chhiber[1,26], S. Dasso[6,24], L. Del Zanna[23,32], P. Dmitruk[6,24], L. Franci[7], M. L. Goldstein[33], A. Greco[4], T. S. Horbury[8], H. Ji[22], J. C. Kasper[9], K. G. Klein[10], S. Landi[23,32], B. Lavraud[30], O. Le Contel[25], H. Li[11], F. Malara[4], B. A. Maruca[1], P. Mininni[6,24], S. Oughton[12], E. Papini[29], T. N. Parashar[1,27], F. Pecora[1], A. Petrosyan[13], A. Pouquet[14], A. Retino[25], O. Roberts[15], D. Ruffolo[16], C. Salem[21], S. Servidio[4], M. Shay[1], H. Spence[17], C. W. Smith[17], J. E. Stawarz[8], M. L. Stevens[31], J. TenBarge[22], B. J. Vasquez[17], A. Vaivads[18], F. Valentini[4], M. Velli[19], A. Verdini[23,32], D. Verscharen[20], R. T. Wicks[28], Y. Yang[1], G. Zimbardo[4]

[1]University of Delaware, Newark DE USA, [2]Swarthmore College, Swarthmore, PA USA, [3]Space Science Institute, Boulder CO USA, [4]Università della Calabria, Rende, Italy, [5]University of Chicago, USA, [6]University of Buenos Aires, Argentina, [7]Queen Mary University of London, UK, [8]Imperial College London, London, UK, [9]University of Michigan, Ann Arbor MI USA, [10]University of Arizona, USA, [11]Los Alamos national Laboratory, Los Alamos NM, USA, [12]University of Waikato, Hamilton NZ, [13]Space Research Institute, Russian Academy of Sciences, Moscow, [14]Laboratory for Atmospheric and Space Physics, University of Colorado Boulder USA, [15]Space Research Institute, Austrian Academy of Sciences, Graz, Austria, [16]Mahidol University, Bangkok, Thailand, [17]University of New Hampshire, Durham NH, USA, [18]KTH, Stokcholm, Sweden, [19]University of California Los Angeles, CA USA, [20]University College London, UK, [21]University of California Berkeley, CA USA, [22]Princeton University, NJ USA, [23]University of Florence, Italy, [24]CONICET, Buenos Aires, Argentina, [25]Ecole Polytechnique, Palaiseau, France; [26] NASA Goddard Space Flight Center, Greenbelt, MD, USA, [27]Victoria University of Wellington, NZ, [28]Northumbria University, Newcastle-upon-Tyne, UK, [29]INAF-Istituto di Astrofisica e Planetologia Spaziali, Rome, Italy, [30]University of Bordeaux, France, [31]Smithsonian Astrophysical Observatory, [32]INAF - Osservatorio Astrofisico di Arcetri, [33]University of Maryland, Baltimore County, Baltimore, MD, USA

(A white paper submitted for the Heliophysics Decadal Survey)

03 September 2022

*Synopsis.*

Space plasmas are three-dimensional dynamic entities. Except under very special circumstances, their structure in space and their behavior in time are not related in any simple way. Therefore, single spacecraft in situ measurements cannot unambiguously unravel the full space-time structure of the heliospheric plasmas of interest – in the inner heliosphere, in the Geospace environment, or the outer heliosphere. This shortcoming leaves numerous central questions incompletely answered. Deficiencies remain in at least two important subjects - Space Weather and fundamental plasma turbulence theory - due to a lack of a more complete understanding of the space-time structure of dynamic plasmas. Only with multispacecraft measurements over suitable spans of spatial separation and temporal duration can these ambiguities be resolved. These characterizations apply to turbulence across a wide range of scales, and also equally well to shocks, flux ropes, magnetic clouds, current sheets, stream interactions, (confined) plasma disruptions, etc. Here, we will describe the basic requirements for resolving space-time structure in general, using "turbulence" as both an example and a principal target of our study. Several types of missions are suggested to resolve space-time structure throughout the Heliosphere.

## 1. Introduction.

Turbulence refers to complex dynamics of fluid and plasma systems when nonlinear effects, such as advection, the Lorentz force, and Ohm's law are stronger than dissipative effects. Dimensionless parameters such as a Reynolds number measure the ratio of the strengths of nonlinearities and dissipation. The usual picture of turbulence begins with a source of large length-scale fluctuations which, through nonlinear processes, transfer energy by cascade mechanisms across the 'inertial range' to the shorter scale lengths of the kinetic range, where the energy is converted into internal energy of the plasma. Small-scale turbulent motions become so disorderly that theory frequently employs statistical descriptions, even if the dynamics is formally deterministic [1]. Revealing the physics of turbulence in the heliosphere (with implications for astrophysical plasmas in general) will require multi-point observations and an array of spacecraft with a range of inter-spacecraft spatial separations [2]. The technology for such missions has been in large part demonstrated by Cluster and MMS, with essential additional capabilities under development for Helioswarm. The multispacecraft missions suggested below, with their diverse scientific goals, are feasible, and their implementation will have an enormous positive impact on heliospheric applications [3], as well as controlled fusion [47,48].

Complex dynamical couplings in turbulence lead to small-scale dissipation of the energy supplied at large scales – a process described as a cascade. Experiments, observations, and numerical simulations all show that analogous descriptions apply to hydrodynamic fluids, magnetofluids (MHD and Hall-MHD), and weakly collisional plasmas. The greatest similarities are found at the larger scales, while space plasmas differ at small scales and high frequencies due to the deficit of collisions and the concomitant emergence of complex kinetic physics. Understanding turbulence requires intensive study of statistical properties for the varying parameters found in nature. We argue that multi-point measurements over a range of scales are required to make significant progress in solar wind physics, which remains the only turbulent space plasma for which such a program is feasible. An array of spatially distributed spacecraft making measurements at moderately high time cadence can provide a wealth of information for space physics applications, including space weather [37], and would be of great importance in more distant plasma venues, from the corona to the interstellar medium.

## 2. *Turbulence effects in the heliosphere.*

The effects of turbulence are intrinsically multi-scale, and the solar wind cascade process spans decades of spatial and temporal scales. One might view the turbulence cascade as a primary way in which cross-scale couplings are enabled, connecting macroscopic and microscopic physics in essential ways. Among the most impactful macroscopic influences of heliospheric turbulence [3] is its possible role as a driver of coronal heating [31] and subsequent solar wind acceleration, which, despite numerous supporting observations, remains to be fully established. Explaining solar wind acceleration is a fundamental goal of missions such as Parker Solar Probe and Solar Orbiter. In the solar wind, extended heating is likely also due to turbulent cascade [4], which operates at different rates in the high cross helicity fast wind, and in the lower cross helicity slow wind [5]. Cross helicity (Alfvénicity, or strong correlation between the velocity and magnetic fields) slows the development of turbulence initially, but eventually expansion [6] and shear [7] or parametric decay (e.g., [45,46]) cause a systematic reduction of this Alfvénic correlation. Similar turbulence effects account for the radial behavior of the Alfvén ratio (or, residual energy), and spectral steepening in Helios data [8]. Accordingly, turbulence also appears to account well for the radial evolution of the (low-frequency) spectral breakpoint that is closely associated with the systematic increase of the correlation scale of the fluctuations. It is noteworthy that these effects are inconsistent with WKB theory of non-interacting waves [3].

All of the above effects are essentially at the larger collective fluid-like scales. Over a range of scales, extending several decades towards the smaller range, theory suggests that the dynamical development of heliospheric turbulence is responsible for the very important observed features of anisotropy [9] and intermittency [10]. See [3] for details.

Another arena in which turbulence is a major player is the transport, scattering, and acceleration of suprathermal and energetic charged particles. In this case, effects such as pitch-angle scattering also operate in a truly cross-scale manner; with solar wind thermal protons resonantly interacting with turbulent fluctuations at the scale of a few hundred kilometers at 1 au, while 1-10 Gev cosmic rays or solar energetic particles (SEPs) resonantly interact with fluctuations at scales of millions of kilometers. Turbulence amplitudes and spectral anisotropy are central in controlling interactions, including resonances, with these energetic particles, e.g. [11].
Given all these demonstrated or anticipated influences, one may reasonably ask at what level do we understand the turbulence that produces these diverse effects in the heliosphere? The answer seems to be that, even with numerous accumulated observational constraints and a reasonable level of progress based on simulation and theory, there are many fundamental questions that remain to be addressed experimentally. Simple, idealized steady-state inertial range phenomenologies can provide motivation for observed spectral slopes, but physical understanding of these diverse cross-scale effects, even in the inertial range, requires deeper, more detailed knowledge and more advanced observations. Beyond inertial range issues, there are questions about dissipation, and intermittency that involve structures and dynamics at sub-proton kinetic scales [12-14]. Fundamental relationships such as the generalized Ohm's law [49] also involve contributions across a wide range of length scales. Due to the cross-scale couplings and cascade mechanisms involved, the kinetic processes are necessarily driven by the cascade from larger energy-containing scales [15]. This poses further observational challenges for understanding dissipative structures and bulk heating in the corona and solar wind.

***Major questions in solar wind turbulence.*** There are numerous outstanding issues about heliospheric turbulence that have not yet been addressed in observations, *in particular, due to a lack of sufficient spatial and temporal resolution.* Without the associated observations, the field cannot realistically advance beyond its current status. A few examples are given here.

***Unraveling correlations & structure in space and time.*** Solar wind researchers are accustomed to employing the Taylor hypothesis, while plasma wave theorists are accustomed to invoking linear dispersion relations. Both of these provide a one-to-one correspondence of variations in space and variations in time. However, in general, spatial and temporal structures are independent entities. For example, the correlation scale is properly defined using single-time multi-point measurements [16]. Unraveling the space-time relationship is a necessary goal in quantifying and distinguishing the effects of turbulence, waves, reconnection, and other phenomena in space plasmas. Revealing how turbulent energy in a space plasma is distributed in space and time requires multi-point measurements.

***Anisotropy of the spectrum at varying scales [9].*** Spectral information relative to preferred directions, e.g., radial and magnetic field directions, is required to validate or refute available theoretical explanations. Purely phenomenological treatments of course do not provide strong conclusions. Anisotropic measurements are required, necessitating simultaneous multi-point measurements that span three-dimensional spatial directions.

***Direct measurement of scale transfer.*** What are the cascade rate and the heating rate? Can turbulence explain observed heating and the origin of the solar wind? The Yaglom-Kolmogorov $3^{rd}$-order laws [17-19] provide a direct evaluation of energy transfer rates at a given scale. The simplest forms require isotropy or some other simple symmetry. Anisotropic forms of the $3^{rd}$-order law in MHD and beyond have been applied using Cluster or MMS at single scales, but understanding cross-scale transfer requires anisotropic measurement at several scales. Simultaneous 3D multi-point measurements are needed to reveal how turbulent energy is transferred anisotropically across scales [50,42]. These exact laws can also be used to unravel in a systematic way higher-order statistics and intermittency in general.

***Higher-order statistics and coherent structures.*** Intermittency or patchiness is an essential feature of turbulent heating and cascade processes. Indeed, in strong turbulence at high Reynolds numbers, most statistical measures of spectral transfer and dissipation are highly non-uniform. The fourth-order (single-time) statistics provide a baseline measurement of intermittency. The sixth-order statistics are a natural measure of the patchiness of energy transfer. These are fundamental but have not been fully characterized and measured in the solar wind, as they must be measured in anisotropic form, due to the strong influence of the large-scale magnetic field and strong gradients, e.g., in stream interaction regions and shear layers [32], as well as in regions of interaction of turbulence with waves [33,34]. Measurement of 3D structure and orientation of coherent structures near the kinetic proton scales is needed to reveal the role of higher order moments in dissipation, thus requiring simultaneous 3D multi-point measurements at two or more spatial scales.

*Anisotropic scale-dependent relaxation times.* Because of the classic "closure problem" (the n-th moment depends on the (n+1)-th), higher order statistics at least up to 4th order contain fundamental information about the dynamics. The single-time statistics are important, but so too are the decay rates of higher-order correlations. For example, the decay time of the 3rd-order correlations controls spectral evolution. In the context of statistical closures [20], the decay times of the triple correlations are identified with scale-dependent Lagrangian correlation times and are usually treated as the local Kolmogorov time scale, because the dominant sweeping timescale does not induce spectral transfer. In plasma, there are additional available time scales, and understanding 3rd and higher-order correlations becomes more complex. Observational constraints, including measurement of anisotropic 4th-order (and higher) moments, are needed to understand this basic physics. Multi-spacecraft measurements are required over a wide range of scales to assess these crucial dynamical time scales.

3. **Key turbulence measurements.**

*A central quantity of interest is the two-point, two-time correlation of a primitive variable* (e.g., a magnetic field component $b_i$.) This *four-dimensional (4D)* space-time correlation may be defined as $R_{ij}(r,\tau) = \langle b_i(x,t) b_j(x+r, t+\tau) \rangle$, where the brackets denote an ensemble average, or a suitable space-time average. The (trace) wave vector spectrum is $S(k) = [\frac{1}{2\pi}]^3 \int d^3r\, R(r,0) e^{ik \cdot r}$, in which the time lag is zero, as well as the Eulerian frequency spectrum $E(\omega) = \frac{1}{2\pi} \int d\tau\, R(0,\tau) e^{i\omega\tau}$, in which the spatial lag is zero. The full space-time (trace) $S(k,\omega)$ spectrum is analogously defined as the Fourier transform of the 4D space-time $R$. This is an analog of a dispersion relation, but without the expectation of a definite relationship between frequency and wavevector. If a nonzero time lag is retained when the spatial transform is carried out, one arrives at the important quantity $S(k,\tau) = S(k)\Gamma(k,\tau)$. This defines the scale-dependent time correlation (in the Eulerian frame) $\Gamma(k,\tau)$, alluded to in the prior section [51]. The space-time correlation also permits direct tests of the Taylor hypothesis. Observational determination of the 2nd-order, two-time, two-point correlation contains much information that we require, but this is not all that is needed to describe interplanetary turbulence.

*A Relation of central importance is the third order law*, which directly measures energy transfer across scales. The contribution of incompressive transfer is given by the Politano-Pouquet law [17]. Hall effect contributions and compressive contributions [21] can be treated additively. With suitable conditions on time stationarity, a pristine inertial range, etc., the relevant incompressive divergence form is $\nabla_s \cdot \langle \delta z_s^\mp |\delta z_s^\pm|^2 \rangle = -4\epsilon^\pm$ for the increments of the Elsässer fields $z^\pm = \Box \pm \Box$ and lag *s*. Integrating over a volume in lag space and employing Gauss's law yields a surface integral that determines the total incompressive transfer of the $z^\pm$ fields across that surface. A suitable multi-spacecraft configuration (say, a regular tetrahedron) enables an approximate evaluation of this transfer [18, 22]. It is also clear that estimates based on the third order law will improve when a sufficient number of lag directions are available which enables directional averaging to be carried out in an appropriate way [40-42]. *Carrying out this multi-spacecraft measurement provides a direct evaluation of scale transfer with no approximations concerning rotational symmetry.* This approach can be supplemented with single spacecraft results using the frozen-in flow (Taylor) hypothesis, or using assumption of isotropy and other rotational symmetry [23]. This approach can reveal potential cascades to both large and small scales [33-35]. It is also possible to account for the presence of large-scale shear [52].

***Fourth-order correlations are also crucial***, as they quantify intermittency, and drive all the important third-order correlations. In MHD the effect of a mean magnetic field appears in the moment hierarchy at the same order as the 4th-order correlations [36]. Together with the mean field, the 3rd- and 4th-order correlations influence the production of spectral anisotropy [9,24], a major issue in plasma cascade and dissipation [14]. Also at 4th-order is the anisotropic *scale-dependent kurtosis,* a quantity that reveals scale-varying intermittency, anisotropy of coherent structures, and incoherent wave activity, as seen in the reference [25] using MMS data.

Much theoretical attention is paid to the inertial and kinetic cascade ranges in plasma turbulence. However, ***the last stages in which collective flow and field energies are converted into microscopic motions or "heat" are crucial for understanding dissipation***. Two quantities of great importance in this regard are the work done on particles of species $\alpha$, that is, $\mathbf{J}_\alpha \cdot \mathbf{E}$ by the electromagnetic (EM) fields, where $\mathbf{E}$ is the total electric field, and the pressure strain interaction $P_{ij}^\alpha \; S_{ij}^\alpha$, where $\mathbf{P}^\alpha$ is the pressure tensor and $S_{ij}^\alpha = \partial_i u_j^\alpha + \partial_j u_i^\alpha$ is the symmetric rate of the strain tensor, each of species $\alpha$ [26, 27]. Even though these quantities are not sign-definite, as viscous dissipation would be, their net (averaged) values are interpreted as the conversion of EM energy into flow energy, and the conversion of energy in the flow into internal energy, respectively. These channels of energy conversion are agnostic regarding specific mechanisms (e.g., reconnection) that may be producing heating, and are therefore crucial diagnostics for understanding the termination of the cascade and the degeneration of collective motions into internal energy. Multi-spacecraft techniques again enter prominently, as the total ***J*** can be evaluated by curlometer techniques, while the rate of strain tensor can be similarly evaluated by differencing the velocities across various spacecraft pairs.

***Summary.*** Statistical quantities are essential to fully understand turbulence, the energy cascade and dissipation in magnetized plasmas. Correlations are expected to be anisotropic and proper analysis requires measurement at several lag scales, perhaps near the ion kinetic scales, or at larger MHD cascade scales. Methods have been developed to extract space-time information from multi-spacecraft datasets, including wave telescope (or k-filtering) [28] and direct methods that rely on ensemble statistics [29]. Some such methods have been successfully applied in plasma laboratory experiments [e.g., 38]. *A quantitative assessment of the Taylor hypothesis would also be provided by these new space-time measurements.*

4. *Mission concepts.*

*Near 1 AU: Turbulence and HelioSwarm.* The recently selected *HelioSwarm* mission consists of nine spacecraft in orbit near 1 AU, providing 36 baselines for two-point measurements. (See Decadal white paper by Klein et al.) This is a breakthrough mission with regard to resolving space/time ambiguity and for the study of fundamental turbulence properties, covering almost two decades of scale from less than 100 km to more than 1000 km. This is the first Heliophysics mission devoted principally to multiscale turbulence physics and will act as a pathfinder for the demonstration of a variety of turbulence properties and analysis techniques. It is expected to reveal fundamental physics of turbulence in the solar wind [30], with immediate implications for the corona [31] and other space and astrophysical plasmas. HelioSwarm targets fundamental science with second and third-order statistics, as well as higher-order multifractal scalings.

Significant steps beyond HelioSwarm will be to employ a larger number of spacecraft as well as improving payload. To obtain the topology of the turbulence and even images of the turbulence, the MagneToRE concept [39], deploys "many magnetometers over 100's of km" to enable the reconstruction of images of the magnetic field using AI techniques. Scintillations of the inter-spacecraft radio communications signals can also provide images of the plasma density. Preliminary analysis shows that approximately 30 nanosats each carrying radio and magnetometer, can accomplish such image reconstruction. This new type of helioscience contributes to Space Weather analysis and allows discovery of the morphology of the interplanetary plasma. Improvements in payload with respect to HelioSwarm will enable improved understanding of energy conversion and particle energization in turbulence. The Plasma Observatory, a multi-scale mission concept submitted to ESA as an M7 candidate, is a constellation of one mothercraft and six identical smallsat daughter craft in 8 $R_E$ ×18 $R_E$ equatorial orbit. A more advanced payload includes electron, mass-resolved ion and energetic particles detectors and electric field antennas. Such a mission would study cross-scale coupling, energization and energy transport within the complex magnetospheric plasmas. MagneToRE and Plasma Observatory are just two examples of the next steps in understanding the 3D structure of the Heliospheric plasma and as the field evolves beyond HelioSwarm.

*L1 Cluster/"MMS in Solar Wind".* The numerous breakthrough discoveries in space plasma physics due to the MMS mission are of enormous significance in magnetospheric reconnection turbulence, and fundamental kinetic plasma physics. However, optimized for magnetospheric goals, its instrument design is less effective in the solar wind [55]. Consequently, there is significant motivation to investigate solar wind physics in similar detail, by deploying a four-spacecraft interplanetary mission having high-resolution instruments based on adaptations of MMS technology. Such a mission would revolutionize solar wind physics.

If a solar wind-adapted MMS-like cluster is placed at L1 in a halo orbit similar to ACE, the mission contributes significantly more to heliospheric physics. Even without continuous high cadence measurements, the L1 cluster would provide considerable support for contemporaneous L1 space weather monitors (such as ACE or IMAP). The L1 cluster provides additional points for estimating structure sizes and gradients that can refine boundary conditions for use in global magnetospheric modeling of response to solar wind conditions. The curvature of field lines, shocks, surfaces of discontinuity, and measurement of the internal structure of the heliospheric current sheet are other major enhancements to L1 monitoring that would be enabled.

*Space Weather Missions in the Inner Heliosphere.* Multispacecraft missions situated well inside of 1 AU can make major contributions to the science underlying space weather as well as contributions to prediction or operations. In this regard, over the past two decades, there have been concept studies such as Solar Wind Sentinels, Inner Heliospheric Mappers, Solar Flotilla, Heliospheric Constellation, and Multispacecraft Heliospheric Mission. These have been discussed as contributions to Space Weather and NASA Living with a Star science and as such have emphasized the detection of the size, shape, and trajectories of relatively large-scale structures such as CMEs, CIRs, shock surfaces, etc. A major goal is usually to improve predictions of solar wind conditions later at Earth. Various formations have been suggested. For example, *Sentinels* employed solar sails (10 m x 100 m) to reduce the solar-directed acceleration on the spacecraft, so that they can orbit the Sun with a one-year period well inside the L1 point at approximately 0.98 AU or climb out of the ecliptic plane to provide information on vertical gradients. These spacecraft would be spaced up to 0.1 AU apart longitudinally bracketing the Sun-Earth line and providing enhanced warnings for corotating solar wind structures. Measurements of the gradients would enable more accurate numerical models, improving predictions of their size and duration of disturbances at Earth. Such a mission might study the physics of shocks and shock particle acceleration, utilize spatial and temporal scales for energy dissipation and transfer in the solar wind, study the magnetic structure of the inner heliosphere, and visualize the dynamics of the inner heliosphere. This kind of *Solar Flotilla* or *Heliospheric Constellation* could consist of multiple autonomous microsatellites in several (three) principal solar elliptic orbits at 0.2 to 0.4 AU, with two to six microsatellites per orbit. Significant variation in inter-spacecraft separation would be possible depending on the orbital details. This class of mission, while its primary goals would be space weather-related, would also make significant contributions to very-large-scale nonlinear plasma dynamics (e.g., origins of turbulence) and the understanding of magnetic structures and SEP propagation.

***Cluster of Solar Probes.*** The Parker Solar Probe Mission has made and will continue to make breakthrough discoveries in the regions of the inner heliosphere that have never been explored before. The unprecedented close approach to the Sun allows the pioneering exploration of exciting features such as: switchbacks in the interplanetary magnetic field; imprints of features of the solar photosphere, chromosphere, and corona in the properties of solar wind fluctuations; radial evolution of solar wind fluctuations, especially crossing the Alfvén critical zone as the wind becomes super-Alfvénic; and a variety of structures in and around the heliospheric current sheet. However, many unsolved problems remain. A future cluster of Parker Solar Probes will be needed to answer questions that a single spacecraft alone cannot resolve. For example, a cluster of probes could: reveal the (local) 3D structure of the Alfvén critical zone [43]; clarify spectral and correlation anisotropy in the young solar wind and its radial evolution [44]; investigate the substructure of the heliospheric current sheet; resolve the topology of newborn magnetic structures; more accurately measure and map the transfer of energy in the lower corona [53,54]; and explore the spatial and temporal structure and intermittency of energy conversion and dissipation processes as a function of distance from the Sun.

## 5. Conclusions.

*Future interplanetary missions with a sufficient number of spacecraft will employ multi-point methods to answer many basic questions about three-dimensional structure and dynamics, related to turbulence and Space Weather, that can be experimentally addressed in no other way [2]. Such missions may span various ranges of scales, varying inter-spacecraft separations depending on their particular emphasis, and may make significant contributions by placing them in various positions in the heliosphere, including near Earth, close to the Sun, and at intermediate distances. These missions are technologically feasible and can be appropriately prioritized to maximize scientific and societal benefits.*